\documentclass[12pt, centerh1]{article}
\usepackage{natbib}
\usepackage[margin=1in]{geometry}
\usepackage{amsmath,natbib}
\usepackage{amsfonts,mathrsfs,moreverb,lipsum}
\usepackage{graphicx,colonequals,multirow}
\usepackage{fixltx2e}
\usepackage{color}
\usepackage{caption}
\usepackage{subcaption}
\usepackage{pbox}
\newcommand{\bs}{\boldsymbol}
\newcommand{\ident}{\mathbf{I}}

\newcommand{\btheta}{{\boldsymbol{\theta}}}
\newcommand{\bvtheta}{{\boldsymbol{\vartheta}}}
\newcommand{\pig}{\pi_g}

\newcommand{\vecmu}{\mbox{\boldmath$\mu$}}

\newcommand{\vecx}{\mathbf{x}}

\newcommand{\vecX}{\mathbf{X}}
\newcommand{\vecU}{\mathbf{U}}

\newcommand{\matsig}{\mathbf\Sigma}

\newcommand{\varthet}{\mbox{\boldmath$\vartheta$}}

\newcommand{\tr}{\,\mbox{tr}}

\newcommand{\vecC}{\mathbf{C}}
\newcommand{\vecsC}{\mathbf{c}}

\newcommand{\vecS}{\mathbf{S}}

\newcommand{\vecalp}{\boldsymbol{\alpha}}

\usepackage{authblk}

\providecommand{\keywords}[1]{\textbf{Keywords}: #1}
\title{Flexible Clustering with a Sparse Mixture of Generalized Hyperbolic Distributions}
\author{Alexa A.\ Sochaniwsky$^1$ \quad Michael B.P.\ Gallaugher\\ \quad \ Yang Tang$^1$\qquad\qquad\quad Paul D.\ McNicholas$^1$}
\date{\small $^1$Dept.\ of Mathematics and Statistics, McMaster University, Hamilton, ON, Canada.\\
$^2$Department of Statistical Science, Baylor University, TX, USA}

\begin{document}
\maketitle{}
\begin{abstract}
Robust clustering of high-dimensional data is an important topic because clusters in real datasets are often heavy-tailed and/or asymmetric. Traditional approaches to model-based clustering often fail for high dimensional data, e.g., due to the number of free covariance parameters. A parametrization of the component scale matrices for the mixture of generalized hyperbolic distributions is proposed. This parameterization includes a penalty term in the likelihood. An analytically feasible expectation-maximization algorithm is developed by placing a gamma-lasso penalty constraining the concentration matrix. The proposed methodology is investigated through simulation studies and illustrated using two real datasets.
\end{abstract}

\section{Introduction}
\begin{figure*}[!htb]
   \centering
   \includegraphics[width=0.95\textwidth]{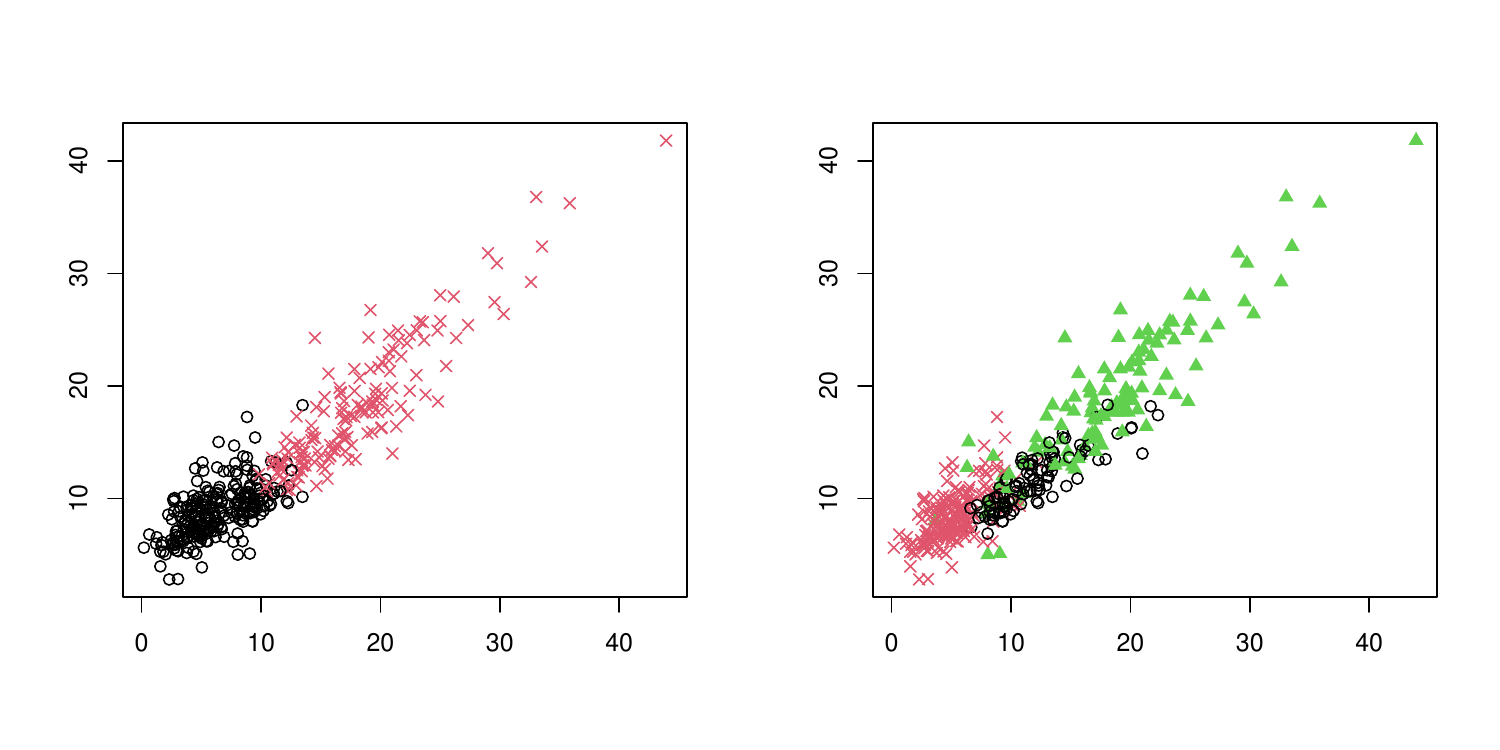}
\vspace{-0.45in}
 \caption{Scatter plots of a two-component mixture of generalized hyperbolic distributions with colour used for true labels (left) and predicted labels from Gaussian model-based clustering (right).}
\label{fig:GHDGMM}
\end{figure*}

In recent years, the use of finite mixture distributions to model heterogeneous data has undergone intensive development in numerous fields such as pattern recognition, cluster analysis, and bioinformatics. Traditionally, Gaussian mixture models dominated the literature; however, when clusters are asymmetric and/or have heavier tails, using Gaussian mixture models tend to overestimate the number of clusters and the result can be clustering results that are not useful in practice \citep[see, e.g.,][]{franczak14}. Consider the data in Fig.~\ref{fig:GHDGMM}, where two asymmetric clusters are generated from a $G=2$ component generalized hyperbolic distribution \citep{browne15}. Gaussian mixtures are fitted to these data for $G=1,\ldots,6$ components and the Bayesian Information criterion \citep[BIC;][]{schwarz78} selects a $G=3$ component model. Notably, the Gaussian components cannot be merged to return the correct clusters (Fig.~\ref{fig:GHDGMM}). For reasons such as this, recent work on model-based clustering has focused on mixtures of non-elliptical distributions \citep[e.g.,][]{murray20,gallaugher22b,dang23,tomarchio23,mclaughlin24}. See \cite{mcnicholas16b} for a review of model-based clustering.
\vspace{-0.45in}

In the case of high dimensional data, traditional methods in the area of asymmetric/flexible model-based clustering and classification can fail due to the number of free covariance parameters and, therefore, alternative techniques are sometimes needed. One such technique is to map the data to a (much) lower dimensional space. Mixtures of factor analyzers models with non-elliptical distributions took off in the last decade, including work on multivariate skew-t distributions \citep{murray14b,murray14a} and generalized hyperbolic distributions \citep{tortora16}. These methods work well with particular datasets; however, considering that this formulation of the skew-t distribution is a special and limiting case of the generalized hyperbolic distribution, they tend to work well on the same datasets. The notion of a joint generalized hyperbolic distribution, which accounts for cluster-specific subspaces, has also been considered \citep{tang18}. 

Various parameterizations of the component covariance matrices $\boldsymbol \Sigma_1, \ldots, \boldsymbol \Sigma_G$ have been considered for dimension reduction in Gaussian mixture models \citep[e.g.,][]{banfield93,celeux95, bouveyron07a}. \cite{krishnamurthy11} consider a sparse covariance matrix for Gaussian mixture models by including a penalty term in the likelihood.
Herein, the work of \cite{krishnamurthy11} is extended by considering a mixture of generalized hyperbolic distributions. The method in \cite{krishnamurthy11} also involves a Laplace prior on each element of the concentration matrix; herein, a gamma hyperprior is considered for the hyperparameter. 

\section{Background}\label{sec:background}
\subsection{Model-Based Clustering}
Model-based clustering, using a finite mixture model, is a common clustering approach. A $G$-component finite mixture model assumes a random vector $\vecX$ has density 
$$
f(\vecx~|~\bvtheta)=\sum_{g=1}^G\pig f_g(\vecx~|~\btheta_g),
$$
where $\bvtheta=\left(\pi_1,\pi_2,\ldots,\pi_G,\btheta_1,\btheta_2,\ldots,\btheta_G\right)$, $f_g(\cdot)$ is the $g$th component density, and $\pig>0$ is the $g$th mixing proportion such that $\sum_{g=1}^G\pig=1$. Typically, the mixture model assumes the component densities to be of the same type for each component $g$, e.g., Gaussian, t-distribution, etc. \cite{mcnicholas16a} traces the relationship between clustering and mixture models all the way back to \cite{tiedeman55}, with the earliest use of a mixture model for clustering presented in \cite{wolfe65} who used a Gaussian mixture model. Other early work in the area of Gaussian mixture models can be found in \cite{baum70} and \cite{scott71}. 

The mathematical tractability of the Gaussian mixture model has made it very popular in the literature; however, a Gaussian distribution may not be appropriate in the presence of asymmetry and/or heavy tails . To combat this issue, work in the area of non-Gaussian mixtures has become popular including distributions with parameterization for concentration such as the $t$ distribution \citep{peel00,andrews11a,andrews11b,andrews12,lin14} and the power exponential distribution \citep{dang15}. Additionally, in the area of robust clustering, one could consider a contaminated approach \citep{punzo16}, a trimmed likelihood approach \citep{escudero20}, or the OCLUST algorithm \citep{clark24}. There has also been significant work in the area of mixtures of skewed distributions such as the skew-$t$ distribution \citep{lin10,vrbik12,vrbik14,lee14,murray14b,murray14a}, the normal-inverse Gaussian distribution \citep{karlis09,ohagan16,fang22}, the generalized hyperbolic distribution \citep{browne15,tortora16,wei19,wei20}, and the skewed power exponential distribution \citep{dang23}. 

While there have been some attempts at incorporating shrinkage into model-based clustering \citep[e.g.,][]{bhattacharya14a,casa20}, approaches for high-dimensional data have generally shied away from shrinkage. The work presented herein is a step towards addressing this gap.

\subsection{Sparse Gaussian Mixture Models}
Assume that a sample of random vectors $\vecX_1,\ldots,\vecX_n$ comes from a $p$-dimensional Gaussian population with $G$ subpopulations such that the $g$th component has mean $\vecmu_g$ and covariance matrix~$\matsig_g$. \cite{krishnamurthy11} uses a penalized observed log-likelihood of the form
\begin{equation*}
\mathcal{L}(\bvtheta)=\sum_{i=1}^n\log\left(\sum_{g=1}^G\pig\phi(\btheta_g)\right)+\sum_{g=1}^G\log\vecC_g,
\end{equation*}
where $\vecC_g=\matsig_g^{-1}$ is the concentration matrix for component $g$ and $\phi(\cdot)$ denotes the multivariate Gaussian density. 
If it is assumed that $C_{gij}\sim\text{Laplace}(0,1/\lambda)$, then the penalty term becomes 
$\sum_{g=1}^G\lambda||\vecC_g||_1$, where $||\cdot||_1$ is the sum of the absolute values of the entries of $\vecC_g$.

Parameter estimation for this model requires the use of the graphical lasso method \citep{friedman08} in conjunction with an expectation-maximization (EM) algorithm \citep{dempster77}. The graphical lasso is a method for the maximization of
$$f(\vecC)=\log \det(\vecC)-\tr(\vecS\vecC)-\rho||\vecC||_1,$$
where $\vecS$ is the empirical covariance matrix and $\rho$ is a tuning parameter. 

\subsection{Generalized Hyperbolic Distribution}
Before introducing the generalized hyperbolic distribution, we briefly discuss the generalized inverse Gaussian distribution. A random variable $W$ follows a generalized inverse Gaussian distribution, denoted by $\text{GIG}(a,b,\gamma)$, if its density function can be written as
\begin{equation*}
f(y|a, b, \gamma)=\frac{\left({a}/{b}\right)^{\frac{\gamma}{2}}w^{\gamma-1}}{2K_{\gamma}(\sqrt{ab})}\exp\left\{-\frac{ay+{b}/{w}}{2}\right\},
\end{equation*}
where $a,b\in\mathbb{R}^+$ and
\begin{equation*}
K_{\gamma}(u)=\frac{1}{2}\int_{0}^{\infty}y^{\gamma-1}\exp\left\{-\frac{u}{2}\left(y+\frac{1}{y}\right)\right\}dy
\end{equation*}
is the modified Bessel function of the third kind with index $\gamma\in\mathbb{R}$. 
Expectations of some functions of a GIG random variable have a mathematically tractable form, e.g.:
\begin{equation}
\mathbb{E}(W)=\sqrt{\frac{b}{a}}\frac{K_{\gamma+1}(\sqrt{ab})}{K_{\gamma}(\sqrt{ab})},
\label{eq:ai}
\end{equation}
\begin{equation}
\mathbb{E}\left({1}/{W}\right)=\sqrt{\frac{a}{b}}\frac{K_{\gamma+1}(\sqrt{ab})}{K_{\gamma}(\sqrt{ab})}-\frac{2\gamma}{b},
\label{eq:bi}
\end{equation}
\begin{equation}
\mathbb{E}(\log W)=\log\left(\sqrt{\frac{b}{a}}\right)+\frac{1}{K_{\gamma}(\sqrt{ab})}\frac{\partial}{\partial \gamma}K_{\gamma}(\sqrt{ab}).
\label{eq:ci}
\end{equation}
In their derivation of the generalized hyperbolic distribution, \cite{browne15} rely on an alternative parameterization of the GIG with density given by
\begin{equation}
g(w|\omega,\eta,\gamma)= \frac{\left({w}/{\eta}\right)^{\gamma-1}}{2\eta K_{\gamma}(\omega)}\exp\left\{-\frac{\omega}{2}\left(\frac{w}{\eta}+\frac{\eta}{w}\right)\right\},
\label{eq:I}
\end{equation}
where $\omega=\sqrt{ab}$ and $\eta=\sqrt{b/a}$. For notational clarity, we will denote the parameterization given in \eqref{eq:I} by $\text{I}(\omega,\eta,\gamma)$.

The generalized hyperbolic distribution in \cite{browne15} arises as a special case of a variance-mean mixture model. This representation assumes that the $p$-dimensional random vector $\vecX$ can be written as
$$\vecX=\vecmu+W\vecalp+\sqrt{W}\vecU,$$
where $\vecmu$ is a location parameter, $\vecalp$ is the skewness, $\vecU\sim\mathcal{N}({\bf 0},\matsig)$ and $W\sim \text{I}(\omega,1,\gamma)$. The resulting density of the generalized hyperbolic distribution is 
\begin{equation*}\begin{split}
f_{\text{GH}}(\vecx|\btheta)&=\exp\left\{(\vecx-\vecmu)'\matsig^{-1}\vecalp\right\}\left[\frac{\omega+\delta(\vecx,\vecmu;\matsig)}{\omega+\vecalp'\matsig^{-1}\vecalp}\right]^{\frac{\gamma-p/2}{2}}\\
&\qquad\qquad\qquad\qquad\qquad\qquad\qquad\qquad\times\frac{K_{\gamma-p/2}(\sqrt{[\omega+\vecalp'\matsig^{-1}\vecalp][\omega+\delta(\vecx,\vecmu;\matsig)]})}{(2\pi)^{p/2}|\matsig|^{1/2}K_{\gamma}(\omega)},\\
\end{split}\end{equation*}
where $\gamma\in\mathbb{R}$ is an index parameter, $\omega>0$ is a concentration parameter, and 
$\delta(\vecx,\vecmu;\matsig)=(\vecx-\vecmu)'\matsig^{-1}(\vecx-\vecmu)$.

\section{Methodology}\label{sec:methodology}
\subsection{Overview}\label{sec:me_overview}
To extend the methodology in \cite{krishnamurthy11}, assume that $\lambda_g\sim \text{Gamma}(s,r)$ and $C_{gij}|\lambda_g\sim\text{Laplace}(0,1/\lambda_g)$ for each $g=1,\ldots,G$, where $\vecC_{g}$ is the inverse of a scale matrix $\matsig_g$ and $C_{gij}$ is the random variable corresponding to the element in the $i$th row and $j$th column of $\vecC_{g}$. The joint density of $\vecC_g$ and $\lambda_g$ is given by
\begin{equation*}\begin{split}
f(\vecC_g,\lambda_g)&=\frac{r^s}{\Gamma(s)}\lambda_g^{s-1}\exp\left\{-r\lambda_g\right\}\prod_{i=1}^p\prod_{j=1}^p\frac{\lambda_g}{2}\exp\left\{-\lambda_g|c_{gij}|\right\}\\
&=\frac{r^s}{\Gamma(s)2^{p^2}}\lambda_g^{s+p^2-1}\exp\left\{-\lambda_g(r+||\vecC_g||_1)\right\},
\end{split}\end{equation*}
and it can be shown that $\lambda_g|\vecsC_g\sim\text{gamma}(s+p^2,r+||\vecC_g||_1)$ and that the marginal distribution of $\vecC_g$ is
$$f(\vecC_g)=\frac{r^s}{\Gamma(s)2^p}\left[\frac{\Gamma(s+p^2)}{(r+||\vecC_g||_1)^{s+p^2}}\right].$$
Note that Stirling's formula \citep{demoivre30} is used to estimate the Gamma function in the marginal distribution.

Suppose we observe a random sample $\vecx_1, \ldots, \vecx_n$ from a $G$-component mixture of generalized hyperbolic distributions. Following \cite{krishnamurthy11}, the observed penalized log-likelihood is
\begin{equation*}\begin{split}
\mathcal{L}(\varthet)=&\sum_{i=1}^N\log \sum_{g=1}^G\pig f_{\text{GH}}(\vecx_i|\vecmu_g,\vecalp_g,\matsig_g,\omega_g,\gamma_g)+\sum_{g=1}^G\log f(\vecsC_{g}).
\end{split}\end{equation*}

\subsection{Parameter Estimation}\label{sec:par.est}
Define $z_{ig}$ so that $z_{ig}=1$ if $\vecx_i$ is in component~$g$ and $z_{ig}=0$ otherwise. Now, the complete-data comprise the $z_{ig}$ together with the latent $w_{ig}$ and the unknown $\lambda_g$ for $i=1,\ldots,n$ and $g=1,\ldots,G$, and the complete-data penalized log-likelihood is
\begin{equation*}
\begin{split}
&\mathcal{L}_C(\varthet)=K+\frac{1}{2}\sum_{i=1}^N\sum_{g=1}^G z_{ig}\log|\vecC_g|
+\sum_{i=1}^N\sum_{g=1}^Gz_{ig}\log h(w_{ig}~|~\omega_g,\gamma_g)\\
&-\frac{1}{2}\tr\bigg\{\sum_{g=1}^G\vecC_g\sum_{i=1}^Nz_{ig}[(1/w_{ig})(\vecx_i-\vecmu_g)(\vecx_i-\vecmu_g)'
-(\vecx_i-\vecmu_g)\vecalp_g'-\vecalp_g(\vecx_i-\vecmu_g)'+w_{ig}\vecalp_g\vecalp_g']\bigg\}\\
&-\sum_{g=1}^G\lambda_g||\vecC_g||_1,
\end{split}
\end{equation*}
where $K$ is a constant with respect to the parameters. An EM algorithm is used to maximize the complete-data likelihood and an outline is given below.

\paragraph{E-step} Update $\hat{z}_{ig}, a_{ig}, b_{ig}, c_{ig}, \hat{\lambda}_g$, where
\begin{align*}
\hat{z}_{ig}&:=\mathbb{E}\left[z_{ig}|\vecx_i\right]=\frac{\hat{\pi}_g^{(t)} f_{\text{GH}}(\vecX_i~|~\hat{\bvtheta}^{(t)}_g)}
{\sum_{h=1}^G\hat{\pi}_h^{(t)}  f_{\text{GH}}(\vecX_i~|~\hat{\bvtheta}^{(t)}_h)},\\
a_{ig}&:=\mathbb{E}\left[W_{ig}~|~\vecx_{i},z_{ig}=1\right],\quad
b_{ig}:=\mathbb{E}\left[{1}/{W_{ig}}|\vecx_{i},z_{ig}=1\right],\quad
c_{ig}:=\mathbb{E}\left[\log W_{ig}|\vecx_{i},z_{ig}=1\right],
\end{align*}
and $\hat{\bvtheta}^{(t)}_g=\left\{\hat{\vecmu}^{(t)}_g,\hat{\vecalp}^{(t)}_g,\hat{\gamma}_g^{(t)},\hat{\omega}_g^{(t)},\hat{\matsig}_{g}^{(t)}\right\}$. 
Fortunately, in the generalized hyperbolic case, 
\begin{equation*}
\begin{split}
&W_{ig}|\vecx_{i},z_{ig}=1
\sim \text{I}(\omega_g+\vecalp_g'\vecC_g\vecalp,\omega_g+\delta(\vecx_i,\vecmu_g|\vecC_g),\gamma_g-p/2),
\end{split}
\end{equation*}
and so $a_{ig}$, $b_{ig}$ and $c_{ig}$ can be calculated using \eqref{eq:ai}--\eqref{eq:ci}.
The update for $\lambda_g$ is
$$
\mathbb{E}\left[\lambda_g|\vecsC_g\right]=\frac{s+p^2}{||\hat{\vecsC}_g^{(t)}||_1+r}\equalscolon\hat{\lambda}_g.
$$
Hereafter, we use the notation $n_g=\sum_{i=1}^n\hat{z}_{ig}$, $\bar{a}_g=1/n_g\sum_{i=1}^n\hat{z}_{ig}a_{ig}$, $\bar{b}_g=1/n_g\sum_{i=1}^n\hat{z}_{ig}b_{ig}$ and $\bar{c}_g=1/n_g\sum_{i=1}^n\hat{z}_{ig}c_{ig}$.

\paragraph{M-step} Update $\pig$, $\vecmu_g$, $\vecalp_g$, $\omega_g$, $\gamma_g$ and $\vecC_g$.
The updates for all these parameters, except $\vecC_g$, are identical to those given in \cite{browne15} and are given by 
\begin{equation*}
\begin{split}
\hat{\pi}_g^{(t+1)}&=\frac{1}{n}\sum_{i=1}^n\hat{z}_{ig}, \quad
\hat{\vecmu}^{(t+1)}_g=\frac{\sum_{i=1}^n\hat{z}_{ig}\vecx_i(\bar{a}_gb_{ig}-1)}{\sum_{i=1}^n\hat{z}_{ig}(\bar{a}_gb_{ig}-1)},\quad
 \hat{\vecalp}^{(t+1)}_g=\frac{\sum_{i=1}^n\hat{z}_{ig}\vecx_i(\bar{b}_{g}-b_{ig})}{\sum_{i=1}^n\hat{z}_{ig}(\bar{a}_gb_{ig}-1)}.
\end{split}
\end{equation*}
The updates for $\gamma_g$ and $\omega_g$ cannot be obtained in closed form and have to be updated using numerical techniques. The details are given in \cite{browne15} and the resulting updates are
\begin{align}
&\hat{\gamma}_g^{(t+1)}=\bar{c}_g\hat{\gamma}_g^{(t)}\left[\left.\frac{\partial}{\partial s}\log(K_{s}(\hat{\omega}_g^{(t)}))\right|_{s=\hat{\gamma}_g^{(t)}}\right]^{-1}, \label{eq:lamup}\\
&\hat{\omega}_g^{(t+1)}=\hat{\omega}_g^{(t)}
-\left[\left.\frac{\partial}{\partial s}q(\hat{\gamma}_g^{(t+1)},s)\right|_{s=\hat{\omega}_g^{(t)}}\right]\left[\left.\frac{\partial^2}{\partial s^2}q(\hat{\gamma}_g^{(t+1)},s)\right|_{s=\hat{\omega}_g^{(t)}}\right]^{-1}, \label{eq:omup}
\end{align}
where the derivative of the Bessel function with respect to the index in \eqref{eq:lamup} is calculated numerically and
\begin{equation*}
\begin{split}
q(\gamma_g,\omega_g)=\sum_{i=1}^Nz_{ig}&\bigg[\log(K_{\gamma_g}(\omega_g))
-\gamma_g\log w_{ig}-\frac{1}{2}\omega_g\left(w_{ig}+\frac{1}{w_{ig}}\right)\bigg].
\end{split}
\end{equation*}
The partial derivatives in \eqref{eq:omup} are described in \cite{browne15} and can be written as
\begin{equation*}\begin{split}
\frac{\partial}{\partial \omega_g}q(\gamma_g,\omega_g)&=\frac{1}{2}\left[R_{\gamma_g}(\omega_g)+R_{-\gamma_g}(\omega_g)-(\bar{a}_g+\bar{b}_g)\right],\\
\frac{\partial^2}{\partial \omega_g^2}q(\gamma_g,\omega_g)&=\frac{1}{2}\left[R_{\gamma_g}(\omega_g)^2-\frac{1+2\gamma_g}{\omega_g}R_{\gamma_g}(\omega_g)-1+
R_{-\gamma_g}(\omega_g)^2-\frac{1-2\gamma_g}{\omega_g}R_{-\gamma_g}(\omega_g)-1\right],
\end{split}
\end{equation*}%
where $R_{\gamma_g}(\omega_g)=K_{\gamma_g+1}(\omega_g)/K_{\gamma_g}(\omega_g)$.
The update for $\matsig_g$ is calculated as follows. Using the graphical lasso method, find
\begin{equation*}
\hat{\vecC}_g^{(t+1)}=\underset{\vecC}{\arg\min} \left\{-\log |\vecC|+\tr(\vecC{\bf S}_g)+\frac{\hat{\lambda}_g}{n_g}||\vecC||_1\right\},
\label{eq:Cupdate}
\end{equation*}
where
\begin{equation*}
\begin{split}
{\bf S}_g=\frac{1}{n_g}&\bigg\{\sum_{i=1}^N\hat{z}_{ig}\left[b_{ig}(\vecx_i-\hat{\vecmu}_g^{(t+1)})(\vecx_i-\hat{\vecmu}_g^{(t+1)})'
-(\vecx_i-\hat{\vecmu}_g^{(t+1)})(\hat{\vecalp}_g^{(t+1)})'\right.\\
&\qquad\qquad\qquad\qquad\qquad\qquad\left.-\hat{\vecalp}_g^{(t+1)}(\vecx_i-\hat{\vecmu}_g^{(t+1)})'+a_{ig}\hat{\vecalp}_g^{(t+1)}(\hat{\vecalp}_g^{(t+1)})'\right]\bigg\}.
\end{split}
\end{equation*}
Then, the update for $\matsig_g$ is
$
\hat{\matsig}_{g}^{(t+1)}=\hat{\vecC}^{(t+1)^{-1}}_g.
$
Despite the use of numerical methods throughout the M-step, the monotonously of the likelihood is preserved. Hereafter, we will refer to this model as the GHD-GLS model.

\subsection{Stopping Rule}
\cite{mcnicholas10a} show that it is possible for the likelihood to ``plateau" and then increase again, and so an EM algorithm may be stopped prematurely if lack of progress in the likelihood is used as the stopping rule. An alternative is to use a stopping rule based on the Aitken acceleration \cite{aitken26}. The Aitken acceleration at iteration $t$ is
$$
a^{(t)}=\frac{l^{(t+1)}-l^{(t)}}{l^{(t)}-l^{(t-1)}},
$$
where $l^{(t)}$ is the observed likelihood at iteration $t$. \cite{bohning94} and \cite{lindsay95} consider the quantity
$$l_{\infty}^{(t+1)}=l^{(t)}+\frac{(l^{(t+1)}-l^{(t)})}{1-a^{(t)}},$$
which is an estimate, at iteration $t+1$, of the observed log-likelihood after many iterations. As in \cite{mcnicholas10a}, we terminate the algorithm when $l_{\infty}^{(k+1)}-l^{(k)}\in(0,\epsilon)$, where $\epsilon$ is small and positive.

\subsection{Model Selection}
In a typical clustering scenario, the number of groups is not known {\it a priori} and, therefore, has to be selected using some criterion. Because the lasso penalty term shrinks the elements of the concentration matrices, there will almost certainly be some elements that are 0 and this needs to be considered when selecting the number of groups.
One example in the literature that demonstrates an approach for dealing with this is lasso-penalized BIC \citep[LPBIC;][]{bhattacharya14a}. The method utilizes a quadratic approximation to the penalty term. However, this cannot be derived here because of the form of $f(\vecC_g)$. Therefore, we propose using the BIC with an effective number of non-zero parameters. This method, however, requires a pre-specified cut-off value for determining which elements can be considered zero. In this paper, we use $10^{-5}$ as this cut-off value. 

\section{Simulation Studies}\label{sec:simulation}
\subsection{Overview}
The performance of the GHD-GHL model is assessed in three ways. Experiment~1 (Section~\ref{sec:sim1}) investigates the sensitivity of the model to different values of the gamma hyperparameters, i.e., $(s,r)$, as well as the effectiveness of BIC in choosing the correct model. Experiment~2 (Section~\ref{sec:sim3}) is designed to assess the proposed sparse modelling approach through different dependency patterns among variables for each component. In Experiment~3 (Section~\ref{sec:sim2}), the proposed model is compared with the parsimonious Gaussian mixture models \citep[PGMM;][]{mcnicholas08,mcnicholas10d} from the \texttt{pgmm} package \citep{pgmm} for the {\sf R} software \citep{R23}, 
 the mixture of generalized hyperbolic distributions (MGHD), and the mixture of generalized hyperbolic factor analyzers \citep[MGHFA;][]{tortora16} --- the {\sf R} package \texttt{mixGHD} \citep{tortora21} is used to implement the latter two approaches. All methods are initialized using $k$-means. When the true classes are known, the performance of the GHD-GLS approach can be assessed using the adjusted Rand index \citep[ARI;][]{hubert85}. The ARI has expected value 0 under random classification and takes the value 1 under perfect class agreement.


\subsection{Experiment 1} \label{sec:sim1}
\begin{figure}[!ht]
\centering
\includegraphics[width = 5.5in]{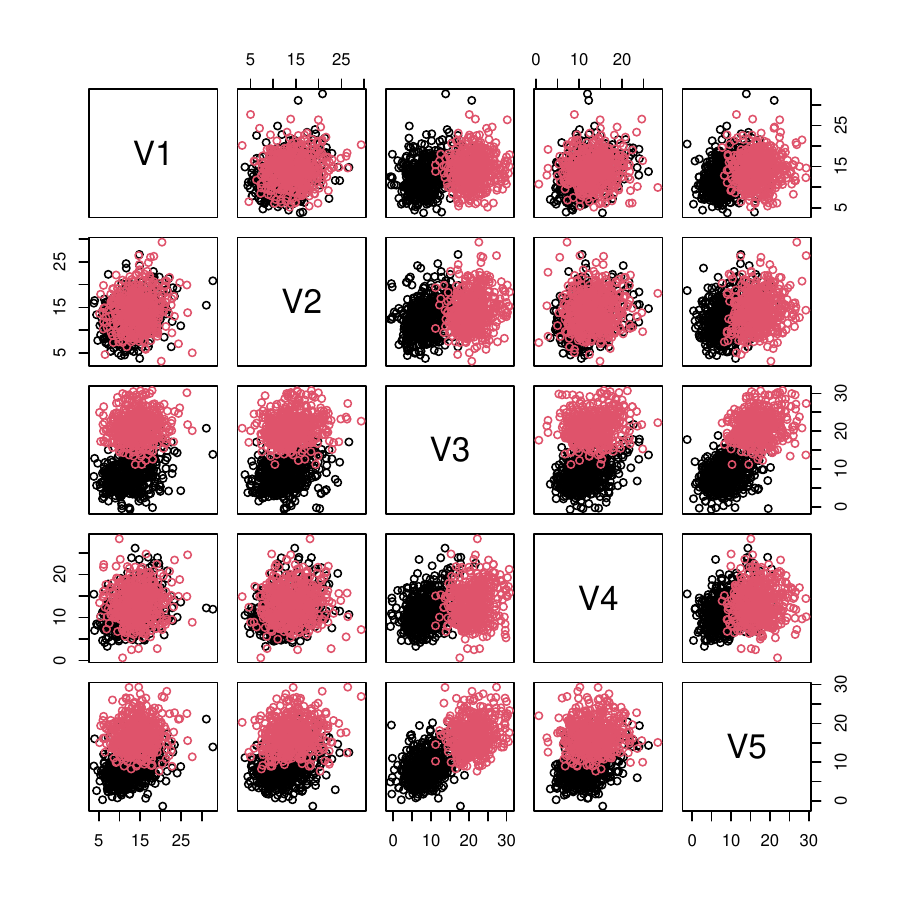}
\caption{The first five variables for one of the simulated datasets from Experiment~1 with $n_g = 500$.}
\label{fig:sim1data}
\end{figure}

A total of 100 samples of each combination of $n_g$ and $(s,r)$ is generated for $G=2$ from a 100 dimension MGHD model (i.e., $p=100$) with $\pi_1=\pi_2 ={1}/{G}$. The scale matrices are of the form $\bs\Sigma_g=\sigma_g^2\ident$ to promote sparsity. An example of the simulated data for $n_g = 500$ is given in Figure \ref{fig:sim1data}, where there is some overlap in the clusters.
Four combinations of the location and rate parameters for the gamma hyperparameters are considered. Table~\ref{table:simsrBIC} shows the BIC and ARI values averaged on the 100 samples for each pair $(s, r)$ as well as the number of times that the correct model is favoured by the BIC for each scenario.
As shown in Table~\ref{table:simsrBIC}, the clustering results do not vary much for different values of the gamma hyperparameters. The BIC usually selects the correct model, and the ARI increases as the number of observations increases. Based on these results, we use $(s,r)=(1,1)$ for the remaining analyses herein. 
\begin{table*}[!ht]
	\centering
	\caption{Values of $n_g$ and $(s,r)$ used in Experiment 1 together with a summary of the results from applying GHD-GLS, i.e., the percent of runs where the BIC selected $G=2$ as well as average BIC and ARI values.}
\begin{tabular*}{1\textwidth}{@{\extracolsep{\fill}}lcccc}
\hline
&$(s,r)$&BIC\%&Avg.\ BIC & Avg.\ ARI\\
\hline
\multirow{3}{*}{$n_g=500$}&$(1,0.5)$& 85 & $-559763.1$ & 0.9954 \\
&$(0.5,0.5)$&85  & $-559774.0$ &0.9954  \\
&$(1,1)$& 88 & $-582372.2$ & 0.9953 \\
&$(0.5,1)$& 87 & $-582374.3$ & 0.9953 \\
\hline
\multirow{3}{*}{$n_g=750$}&$(1,0.5)$& 62  & $-849145.1$ &   0.9959\\
&$(0.5,0.5)$& 59 & $-849144.4$ & 0.9959  \\
&$(1,1)$& 79 & $-859274.5$ & 0.9953 \\
&$(0.5,1)$& 83 & $-859274.1$ & 0.9953 \\
\hline
\multirow{3}{*}{$n_g=1000$}&$(1,0.5)$&100  &$-1122732$  & 0.9942 \\
&$(0.5,0.5)$& 100 & $-1122737$ &  0.9941\\
&$(1,1)$& 100 & $-1127642$ & 0.9936  \\
&$(0.5,1)$& 100 & $-1127641$ & 0.9936 \\
\hline
	\end{tabular*}
	\label{table:simsrBIC}
\end{table*}
%

\subsection{Experiment 2}\label{sec:sim3}
\begin{figure*}[!ht]
    \centering
    \begin{subfigure}[b]{\textwidth}
   \includegraphics[width=0.9\textwidth]{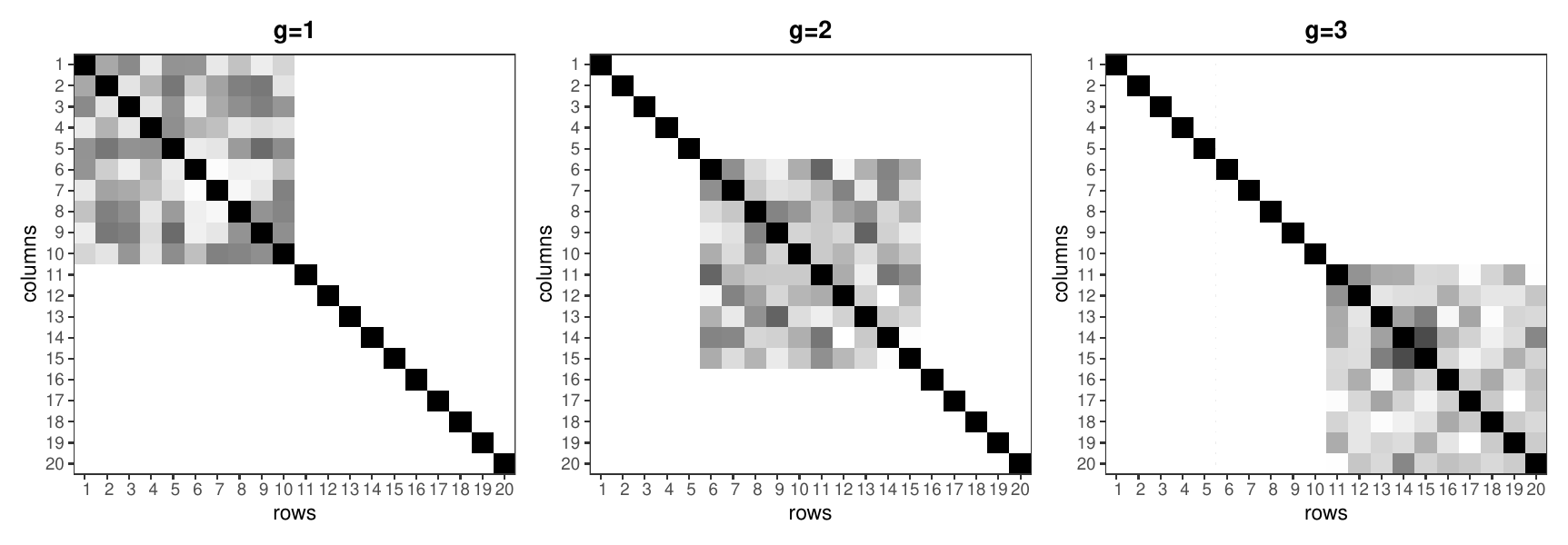}
        \caption{Scenario 1}
        \label{fig:truecov1}
    \end{subfigure}
    ~ 
    \begin{subfigure}[b]{\textwidth}
       \includegraphics[width=0.9\textwidth]{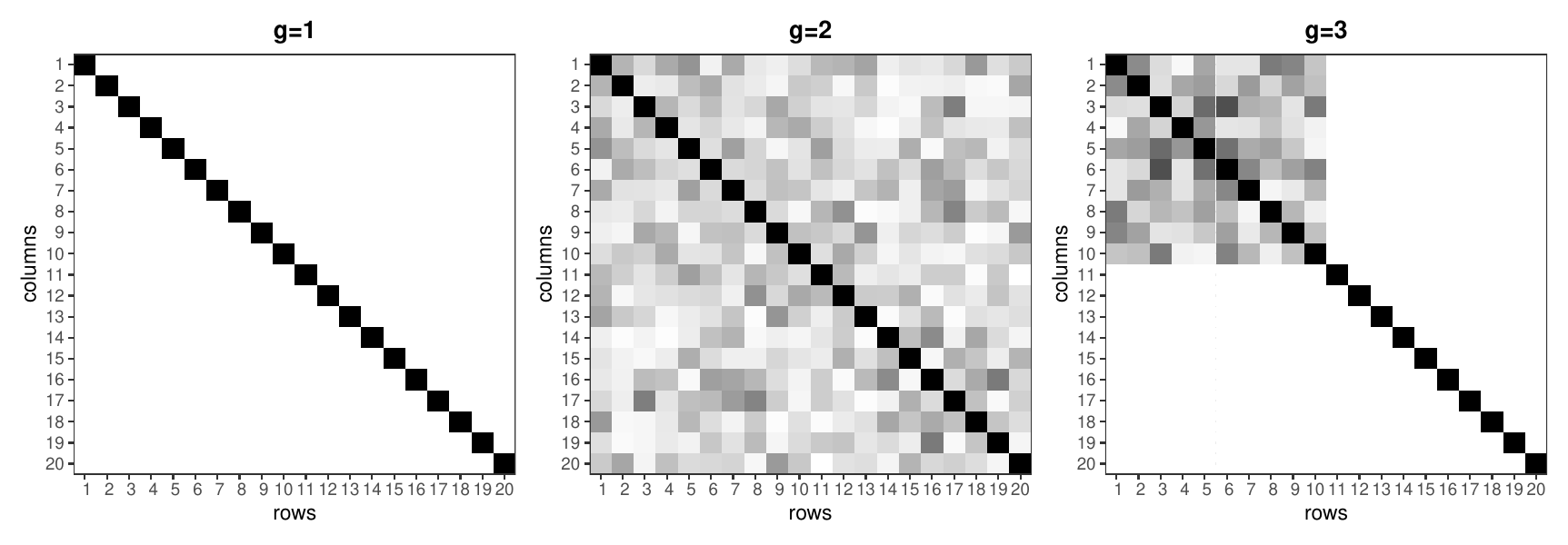}
        \caption{Scenario 2}
        \label{fig:truecov2}
    \end{subfigure}
    \caption{Heat maps of the true component covariance matrices in the two scenarios from Experiment~2.}\label{fig:sim2}
\end{figure*}

\begin{figure*}[!ht]
    \centering
    \begin{subfigure}[b]{1\textwidth}
        \includegraphics[width=1\textwidth]{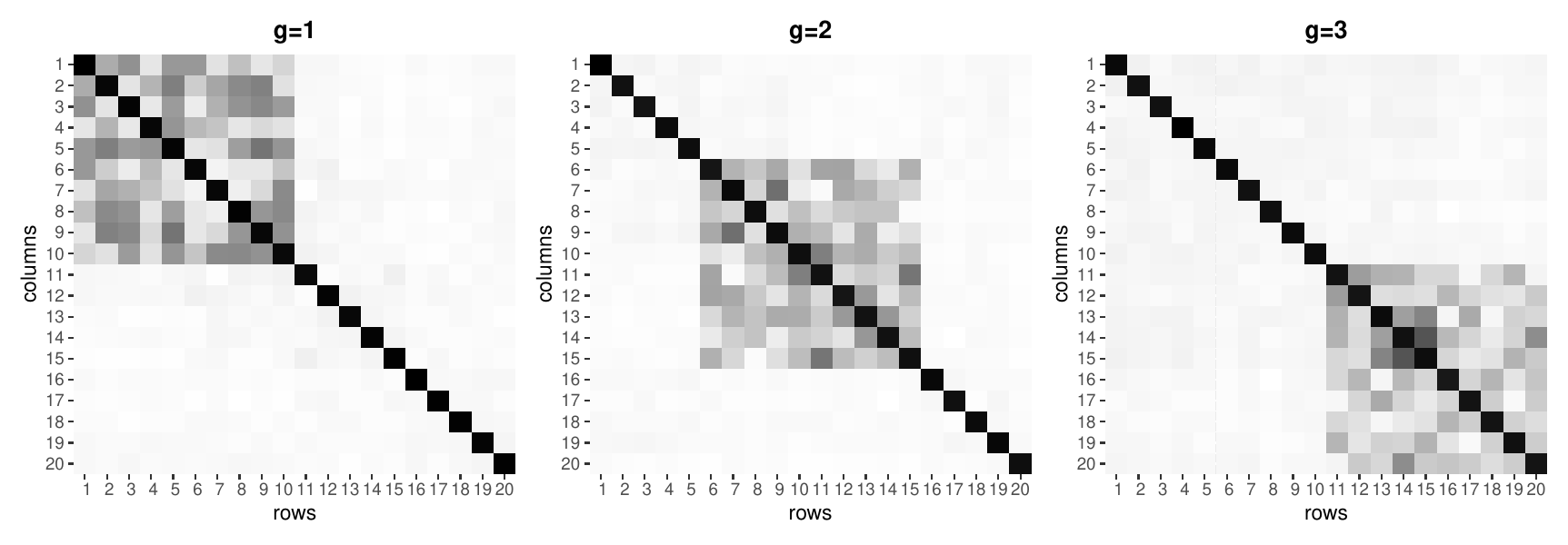}
        \caption{$n_g=250$}
        \label{fig:s1ng50}
    \end{subfigure}
    ~ 
    \begin{subfigure}[b]{1\textwidth}
        \includegraphics[width=1\textwidth]{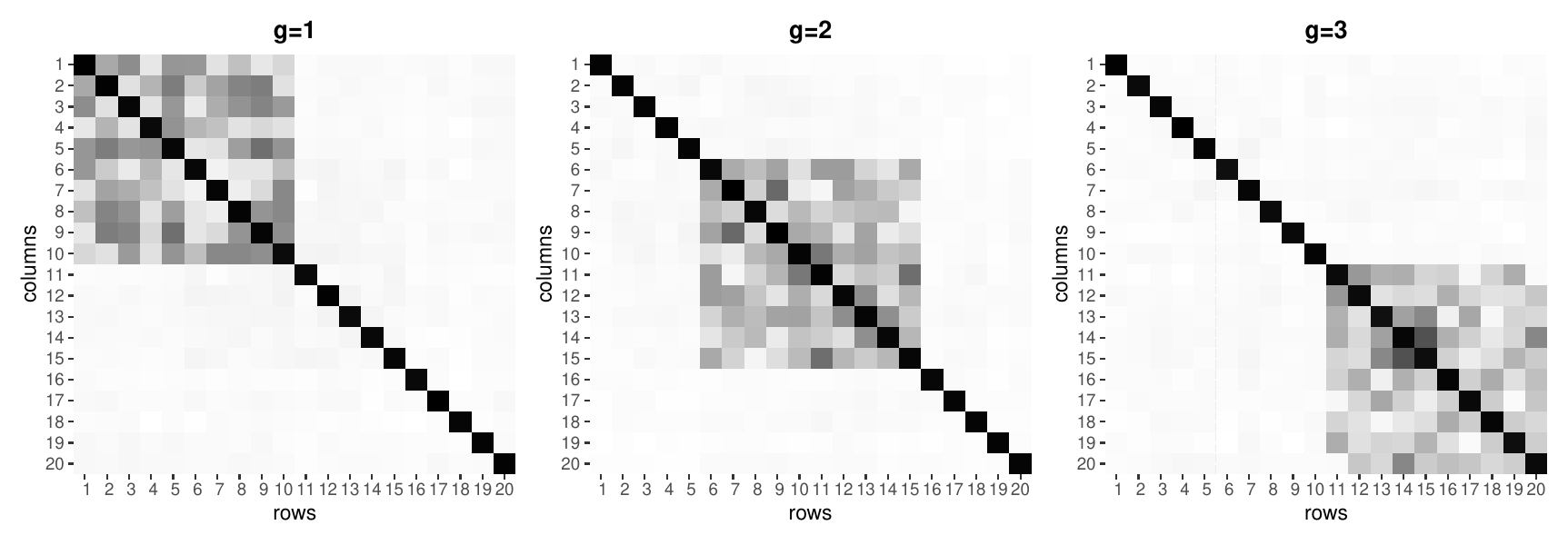}
        \caption{$n_g=500$}
        \label{fig:s1ng100}
    \end{subfigure}
     ~ 
    \begin{subfigure}[b]{1\textwidth}
        \includegraphics[width=1\textwidth]{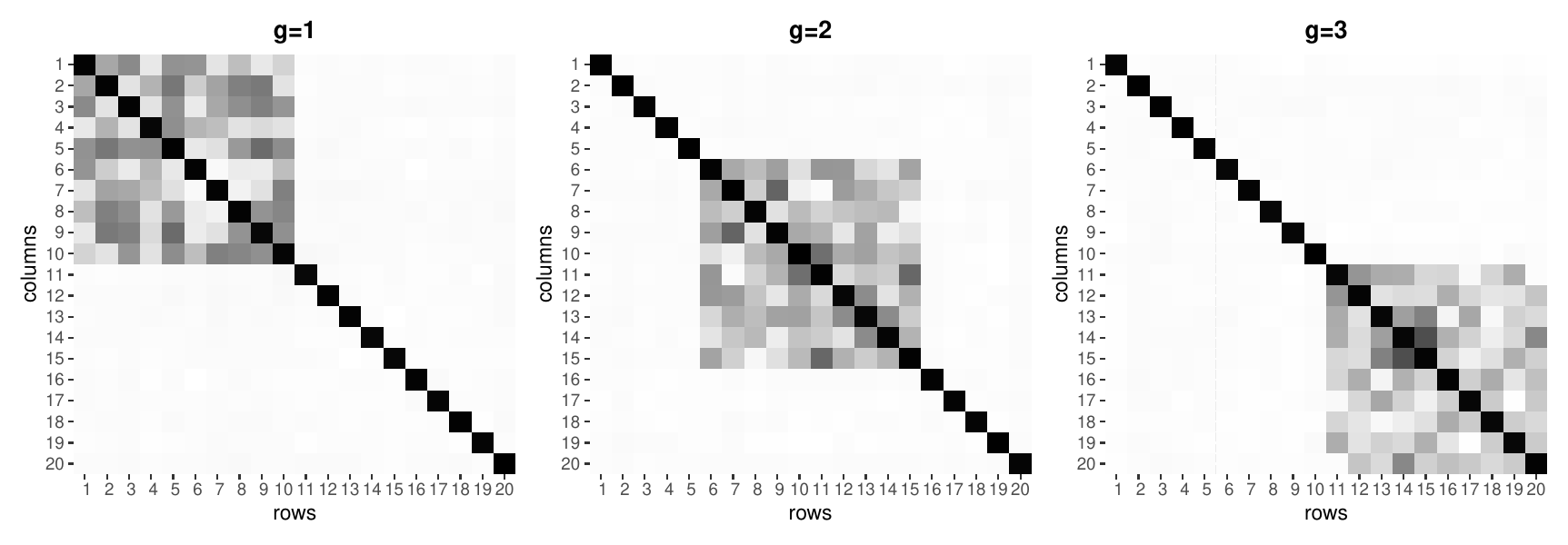}
        \caption{$n_g=1000$}
        \label{fig:s1ng150}
    \end{subfigure}
    \caption{Heat maps of the estimated component covariance matrices in Scenario 1 for each sample size.}\label{fig:sim2results1}
\end{figure*}

\begin{figure*}[!ht]
    \centering
    \begin{subfigure}[b]{1\textwidth}
        \includegraphics[width=1\textwidth]{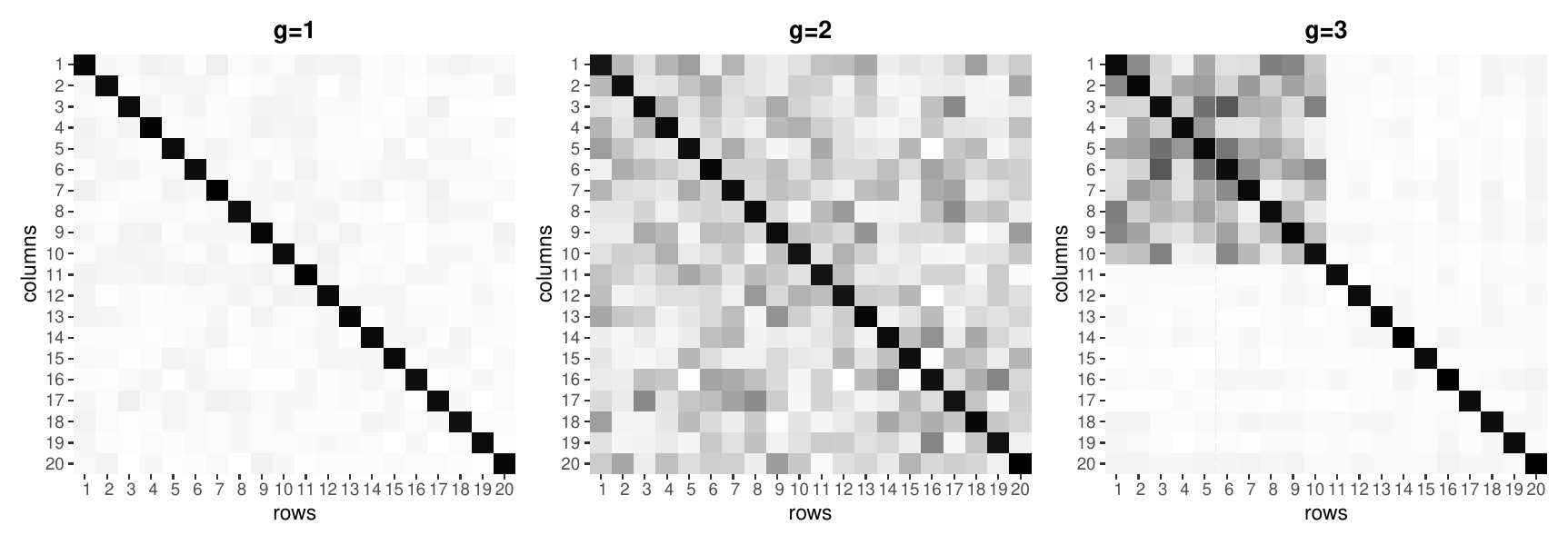}
        \caption{$n_g=250$}
        \label{fig:s2ng50}
    \end{subfigure}
    ~ 
    \begin{subfigure}[b]{1\textwidth}
        \includegraphics[width=1\textwidth]{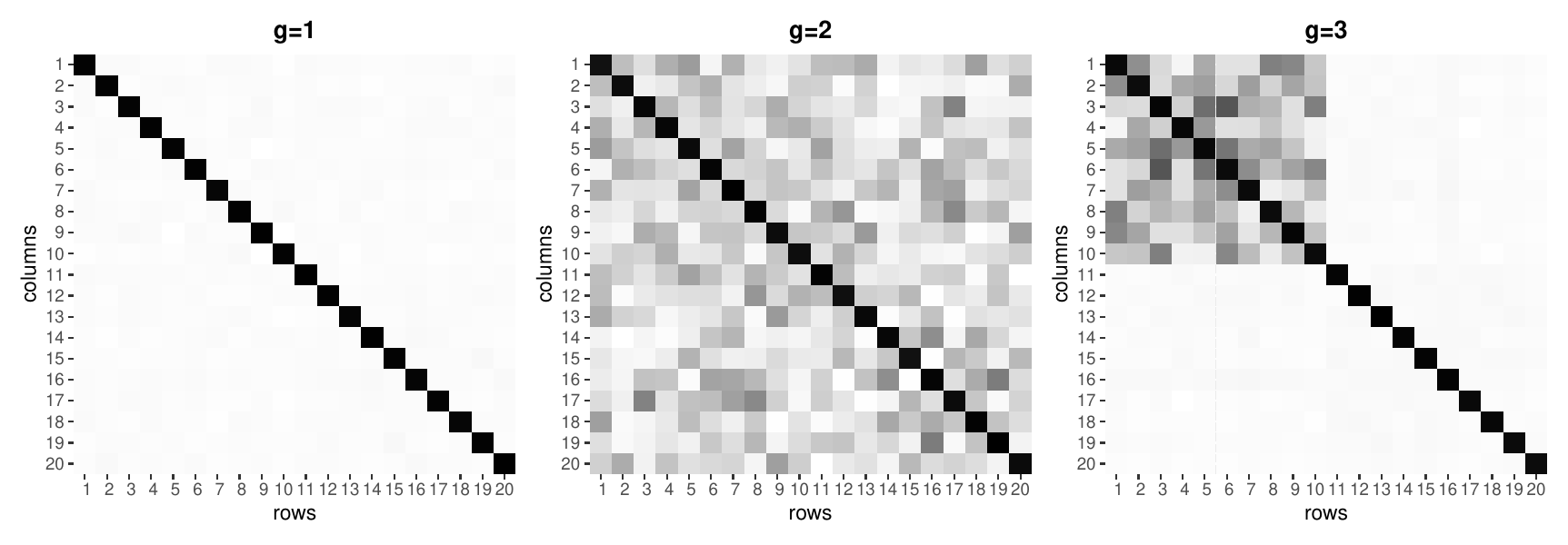}
        \caption{$n_g=500$}
        \label{fig:s2ng100}
    \end{subfigure}
     ~ 
    \begin{subfigure}[b]{1\textwidth}
        \includegraphics[width=1\textwidth]{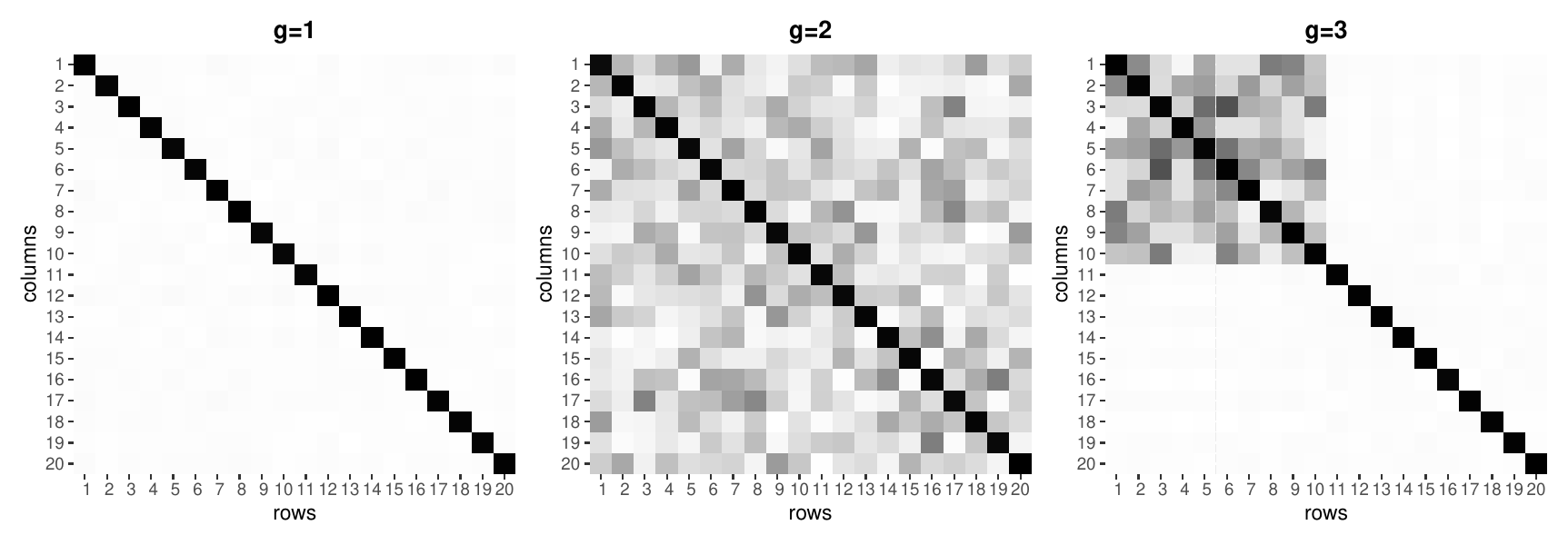}
        \caption{$n_g=1000$}
        \label{fig:s2ng150}
    \end{subfigure}
    \caption{Heat maps of the estimated component covariance matrices in Scenario 2 for each sample size.}\label{fig:sim2results2}
\end{figure*}

Two scenarios are considered with different structures for $\bs\Sigma_g$ and $p=20$ dimensions. Fig.~\ref{fig:sim2} presents the heat maps of the true component covariance structures for the two scenarios. In each scenario, a set of 100 samples for each $n_g\in\{250, 500, 1000\}$ is generated from a three-component MGHD with the corresponding covariance structures. The GHD-GLS models are fitted for $G=3$ with $k$-means starts. Figs.~\ref{fig:sim2results1} and~\ref{fig:sim2results2} show the averaged estimated component covariance matrices for each sample size in Scenarios~1 and~2, respectively. Overall, the GHD-GLS models show promising performance in recovering the underlying structures of the component covariance matrices in both scenarios. As one would expect, the estimation becomes more accurate as the sample size grows.  

\subsection{Experiment 3}\label{sec:sim2}
In this experiment, we compare our proposed approach with three comparator approaches: PGMM, MGHD, and MGHFA. PGMM is developed for high-dimensional symmetrical data whereas the MGHD and MGHFA approaches can model clusters with skewness and/or heavy tails. Two scenarios are considered: one where 100 samples are generated from the MGHD (Scenario 3) and another where 100 samples are generated from a mixture of Gaussian distributions (Scenario 4). Each combination of $n_g\in\{500,750,1000\}$ and $p\in\{20,50,100\}$ is considered for a three-component mixture model with the covariance structure in Fig~\ref{fig:sim2}(a).

Tables~\ref{table:compare1} and \ref{table:compare2} show the averaged BIC and ARI over the 100 samples in Scenarios~3 and~4, respectively. 
Tables~\ref{table:compare3} and \ref{table:compare4} (Appendix~\ref{app:time}) show the average and standard deviation of the times that correspond to the results in Table~\ref{table:compare1} and \ref{table:compare2}, respectively. In Scenario~3, GHD-GLS performs the best overall but MGHD is the fastest.  PGMM and MGHFA are effective for dimension reduction; however, the classification results are significantly worse than for GHD-GLS. In Scenario 4, unsurprisingly, PGMM performs the best among all four approaches because the samples are generated from Gaussian mixtures.
\begin{table*}[!ht]
	\centering
 \small
	\caption{A comparison of the selected GHD-GLS, PGMM, MGHD and MGHFA in Scenario~3 (replications=100).}	
	\scalebox{1}{
		\begin{tabular*}{1.05\textwidth}{@{\extracolsep{\fill}}lccccccccccc}
		  \noalign{\smallskip}\hline
 		&&\multicolumn{2}{c}{GHD-GLS} &\multicolumn{2}{c}{PGMM} &\multicolumn{2}{c}{MGHD} & \multicolumn{2}{c}{MGHFA} \\ 
\cline{3-4}\cline{5-6}\cline{7-8}\cline{9-10}\cline{11-12}
		               &&BIC&ARI&BIC&ARI&BIC&ARI&BIC&ARI\\
 		 \hline
\multirow{3}{*}{$p=20$}&$n_g=500$ 
& $-190781.9$ & 0.870 &  $-186772.5$ & 0.558  &  $-187942.0$ & 0.867 & $-186891.0$ & 0.617 \\
&$n_g=750$& $-283592.8$ & 0.885 & $-279930.3$ & 0.558  & $-280032.0$  & 0.883  & $-279634.1$ &  0.621 \\
&$n_g=1000$& $-376203.6$ &  0.890 &$-373013.5$  &  0.574 & $-372002.3$ & 0.888 & $-372295.0$ & 0.622 \\
\hline
\multirow{3}{*}{$p=50$}&$n_g=500$ 
& $-499851.4$ & 0.926 &$-471148.5$  & 0.551  & $-487045.2$  & 0.915  & $-472169.9$ & 0.534 \\
&$n_g=750$& $-731673.2$ & 0.939 & $-706102.2$ &  0.550  & $-719751.8$ & 0.934  & $-706460.8$ &  0.526 \\
&$n_g=1000$& $-964276.9$ & 0.950 & $-941169.3$  & 0.550  & $-952085.3$ & 0.947 & $-940735.6$ &  0.516 \\
\hline
\multirow{3}{*}{$p=100$}&$n_g=500$ 
& $-1095283$ & 0.952 & $-938231.1$ &  0.530 & $-1017441$  & 0.859 & $-940535.7$ & 0.625 \\
&$n_g=750$& $-1572922$ & 0.973 & $-1406088$ &  0.530 & $-1483629$ & 0.944 &$-1407180$ & 0.604 \\
&$n_g=1000$& $-2045354$ & 0.985 & $-1873921$ & 0.531  & $-1948150$ & 0.972 & $-1873840$ & 0.609 \\
\hline
	\end{tabular*}}
	\label{table:compare1}
\end{table*}
\begin{table*}[!ht]
	\centering
 \small
	\caption{A comparison of the selected GHD-GLS, PGMM, MGHD and MGHFA in Scenario~4 (replications=100).}	
	\scalebox{1}{
		\begin{tabular*}{1.05\textwidth}{@{\extracolsep{\fill}}lccccccccccc}
		  \noalign{\smallskip}\hline
 		&&\multicolumn{2}{c}{GHD-GLS} &\multicolumn{2}{c}{PGMM} &\multicolumn{2}{c}{MGHD} & \multicolumn{2}{c}{MGHFA} \\ 
\cline{3-4}\cline{5-6}\cline{7-8}\cline{9-10}\cline{11-12}
		               &&BIC&ARI&BIC&ARI&BIC&ARI&BIC&ARI\\
 		 \hline
\multirow{3}{*}{$p=20$}&$n_g=500$ 
&  $-170483.1$ & 0.851 & $-164229.7$ & 1 & $-167143.2$ & 0.852 & $-165129.1$  &  0.994 \\
&$n_g=750$& $-253281.6$ & 0.823 & $-245958.0$ & 1 & $-249065.9$ & 0.825 & $-246998.7$ & 0.990 \\
&$n_g=1000$& $-335039.2$ &0.867 & $-327611.4$ & 1& $-330268.1$ & 0.869  & $-328582.3$ & 1 \\
\hline
\multirow{3}{*}{$p=50$}&$n_g=500$& $-450749.6$ & 0.830 & $-417357.2$ &  1 & $-437503.0$ & 0.817  & $-419967.3$ & 0.990 \\
&$n_g=750$& $-664606.7$ & 0.782 &$-625218.5$  & 1 &  $-646100.1$ & 0.772 & $-628190.7$ & 0.978 \\
&$n_g=1000$& $-876314.8$ & 0.757 & $-833059.7$ &  1 & $-854186.0$ & 0.750  & $-835963.5$  & 0.989 \\
\hline
\multirow{3}{*}{$p=100$}&$n_g=500$& $-982765.7$ & 0.819 & $-832882.6$ & 1 & $-919609.7$ & 0.789 & $-837742.0$ & 0.995 \\
&$n_g=750$& $-1449128$ & 0.829 & $-1247655$ & 1 & $-1337731$ & 0.817 & $-1253036$ & 1 \\
&$n_g=1000$& $-1882055$ & 0.806 & $-1662259$ & 1 & $-1754512$ & 0.795 & $-1667926$ & 1 \\
\hline
	\end{tabular*}}
	\label{table:compare2}
\end{table*}

\clearpage

\section{Real Data Analyses} \label{sec:app}
\subsection{Overview}
The proposed GHD-GLS approach is compared with two methods based on generalized hyperbolic distributions, i.e., MGHD and MGHFA, on real data. The Movehub quality of life data (Section~\ref{sec:dataapp1}) is selected to demonstrate the interpretability of our proposed approach, and the breast cancer diagnostic data (Section~\ref{app:data2}) is used because of its popularity as a benchmark dataset within the literature.

\subsection{Movehub Quality of Life} \label{sec:dataapp1}
The Movehub quality of life data consist of five key metrics for 216 cities (i.e., $n=216$ and $p=5$): purchase power, health care, pollution, quality of life, and crime rate. An overall rating for a city is given considering all five metrics. The data are available online at {\tt www.movehub.com}. The GHD-GLS, MGHD, and MGHFA approaches are fitted to these data for $G=2,3,4$ and, for MGHFA, for $q=1,2,3$. The minimum BIC occurs at $G=2$ for MGHD and MGHFA, and at $G=3$ for GHD-GLS. A cross-tabulation of the predicted classifications against true classes for the three approaches is shown in Table~\ref{table:cityclass}. Table~\ref{table:citystats} presents the mean and standard deviation of the five metrics as well as the overall rating for each group. 
\begin{table}[!ht]
	\centering
	\caption{Cross-tabulation of the classes for the selected GHD-GLS model against classes for the selected MGHD and MGHFA models, respectively, for the quality of life data.}	
		\begin{tabular*}{0.75\textwidth}{@{\extracolsep{\fill}}lccccccc}
		 \noalign{\smallskip} \hline
 		GHD-GLS  & \multicolumn{2}{c}{MGHD} && \multicolumn{2}{c}{MGHFA} \\ 
 		\hline
		             &Group 1&Group 2&&Group 1&Group 2\\
\cline{2-3} \cline{5-6}
		Group 1&   84&4&&    77&11\\
		Group 2&   1&83&&   0&84\\
        Group 3&   44&0&&   34&10\\
 		 \hline
	\end{tabular*}
	\label{table:cityclass}
\end{table}
\begin{table*}[!ht]
	\centering
	\caption{Mean and standard deviation of the five key metrics for each component of the three approaches.}
	\scalebox{0.95}{
		\begin{tabular*}{1\textwidth}{@{\extracolsep{\fill}}lcccccc}
		\noalign{\smallskip}  \hline
		 \multicolumn{7}{c}{Group 1} \\
		\cline{2-7}
 		&\multicolumn{2}{c}{GHD-GLS} & \multicolumn{2}{c}{MGHD} & \multicolumn{2}{c}{MGHFA} \\                        
		              &Mean&Std. dev.&Mean&Std. dev.&Mean&Std. dev.\\
 		 \cline{2-3}\cline{4-5}\cline{6-7}
Overall Rating& 83.45  & 4.01 & 83.90 & 3.92 &83.86  & 3.50\\
Purchase Power& 58.14 & 12.51 & 60.90 & 12.08 & 62.70 & 11.92\\
Health Care& 71.75 & 11.07 & 70.94 & 11.36 & 71.85 & 10.14\\
Pollution& 23.07 & 10.10 & 38.48 & 23.89 & 36.57  & 23.20\\
Quality of Life& 76.13 & 9.61 & 74.85 & 11.09  & 77.57 & 8.47\\
Crime Rate& 35.21 & 14.37 & 38.55 & 16.46 & 36.53 & 15.37\\
\hline
 \multicolumn{7}{c}{Group 2} \\
		\cline{2-7}
		&\multicolumn{2}{c}{GHD-GLS} & \multicolumn{2}{c}{MGHD} & \multicolumn{2}{c}{MGHFA} \\                        
		              &Mean&Std. dev.&Mean&Std. dev.&Mean&Std. dev.\\
 		 \cline{2-3}\cline{4-5}\cline{6-7}
Overall Rating& 73.39 & 4.20 & 73.41  & 4.04 & 75.25 & 6.00\\
Purchase Power& 24.95 & 8.67 & 25.10 & 8.42 & 29.33 & 12.26\\
Health Care& 59.52 &16.10 & 59.76 & 15.88 & 60.72 & 16.02\\
Pollution& 56.82 & 23.87 & 55.26 & 24.28 & 54.40 &24.42 \\
Quality of Life& 37.05 & 13.81 & 37.97 &  14.41 & 41.41 &15.75 \\
Crime Rate& 45.88 & 15.51 &  45.48 & 15.54 & 46.43 & 16.01\\
\hline	
\multicolumn{7}{c}{Group 3} \\
		\cline{2-7}
		&\multicolumn{2}{c}{GHD-GLS} & \multicolumn{2}{c}{MGHD} & \multicolumn{2}{c}{MGHFA} \\                        
		              &Mean&Std. dev.&Mean&Std. dev.&Mean&Std. dev.\\
 		 \cline{2-3}\cline{4-5}\cline{6-7}
Overall Rating& 84.13 & 4.27 &NA&NA&NA&NA\\
Purchase Power&  64.25 & 12.53 &NA&NA&NA&NA\\
Health Care& 69.03 &  11.53 &NA&NA&NA&NA\\
Pollution& 67.47 & 11.76 &NA&NA&NA&NA\\
Quality of Life& 71.52 & 13.12 &NA&NA&NA&NA\\
Crime Rate& 44.93 & 18.31 &NA&NA&NA&NA\\	
\hline	
	\end{tabular*}}
	\label{table:citystats}
\end{table*}

\begin{figure*}[!ht]
    \centering
    \begin{subfigure}[b]{3in}
        \includegraphics[width=3in]{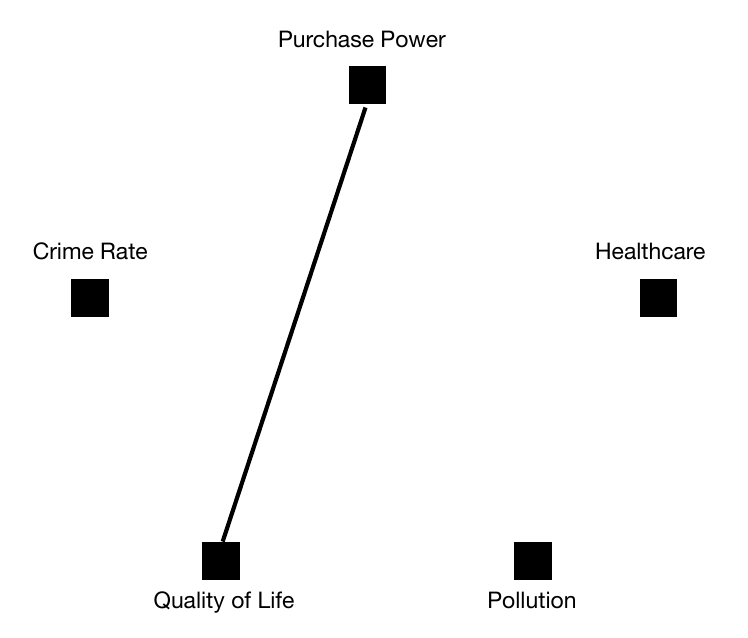}
        \caption{Group 1}
        \label{fig:cityg1}
    \end{subfigure}
    ~ 
    \begin{subfigure}[b]{3in}
        \includegraphics[width=3in]{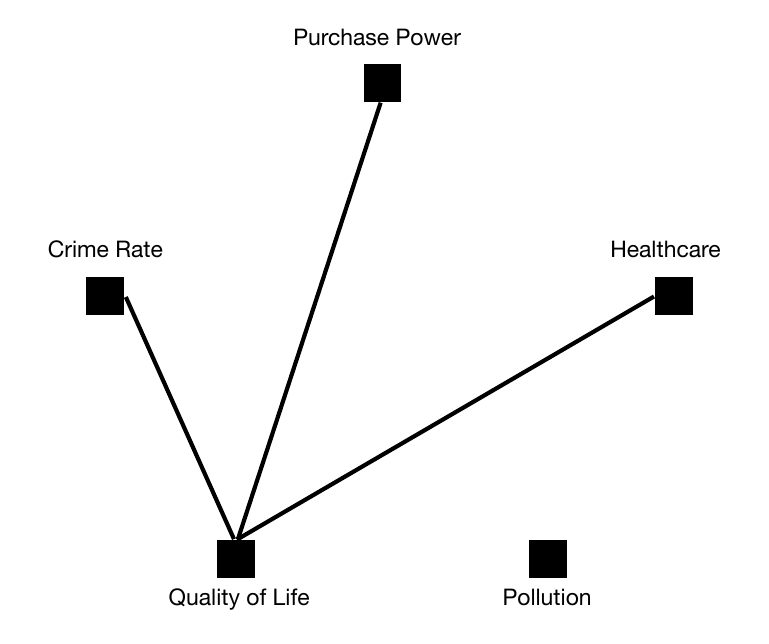}
        \caption{Group 2}
        \label{fig:cityg2}
    \end{subfigure}
     ~ 
    \begin{subfigure}[b]{3in}
        \includegraphics[width=3in]{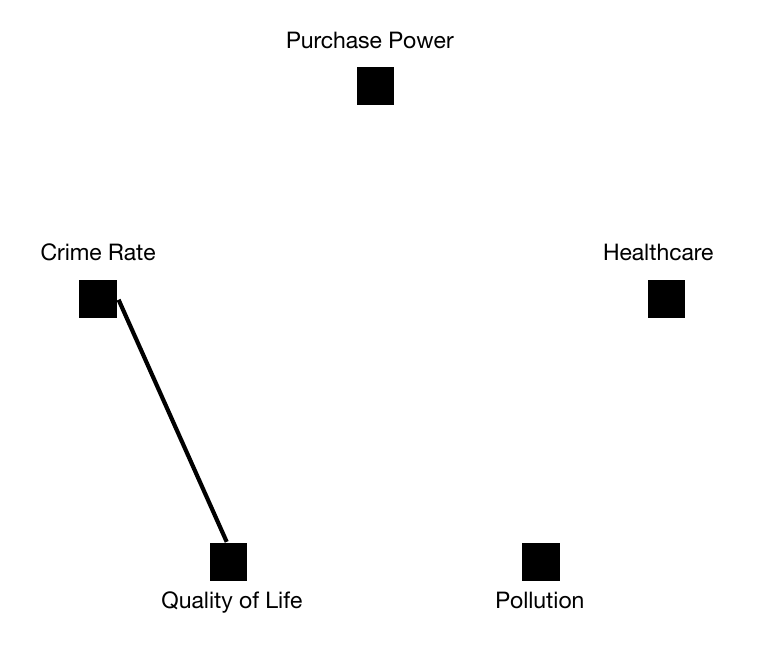}
        \caption{Group 3}
        \label{fig:cityg3}
    \end{subfigure}
    \caption{The correlation structures among the five variables for each group for the quality of life data using the GHD-GLS approach.}\label{fig:cityg}
\end{figure*}

Group 1 consists of cities with lower ratings in purchase power, health care and quality of life, while having higher ratings in pollution and crime rate. The overall rating for the cities in Group 1 is lower when compared to cities in Groups 2 and~3. Fig.~\ref{fig:cityg} shows the sparse correlation structures among the five variables differ across groups which can only be found using the GHD-GLS approach. In particular, Group 1 is characterized by the relation between quality of life and purchase power, Group 2 is characterized by the relation between quality of life, crime rate, health care and purchase power, and Group 3 is characterized by the relation between quality of life and crime rate.

The predicted classification from our approach agrees with MGHD on 167 cities and disagrees on 49 cities. Out of those 49 cities, GHD-GLS distinguishes 44 of those cities as a third group. Similarly, the predicted classification from our approach agrees with MGHFA on 161 cities and disagrees on 55 cities where GHD-GLS distinguishes 34 of those cities as a third group. The remaining 21 cities are placed into Group 2 where they belong to Group 1 or~3 using GHD-GLS.

\subsection{Breast Cancer Diagnostic Dataset} \label{app:data2}
The breast cancer diagnostic data was first used in \citet{street93}. Ten real-valued features on 569 cases of breast tumours are reported: 357 benign and 212 malignant. The mean, standard error, and ``worst'' or largest of these features were computed for each image, resulting in 30 attributes. The GHD-GLS, MGHD and MGHFA methods are fitted for $G=2,3,4$ and, for MGHFA, for $q = 1,2,3$. A summary of the best models from the GHD-GLS, MGHD and MGHFA approaches is given in Table~\ref{table:data2comp.}. The GHD-GLS and MGHD  approaches give the correct number of components, i.e., $G=2$. The GHD-GLS approach yields the best ARI among the three approaches ($\text{ARI}=0.77$) and misclassifies only 34 out of 569 observations (Table~\ref{table:result2}). 
\begin{table}[!ht]
\centering
\vspace{-0.1cm}
\caption{A comparison of the selected GHD-GLS and two different approaches on the tumour data.}
\scalebox{1}{
\begin{tabular*}{0.5\textwidth}{@{\extracolsep{\fill}}lrrr}
\noalign{\smallskip}\hline
&$G$&BIC&ARI\\ [0.5ex] 
\hline
GHD-GLS&$2$&$-36997.49$&$0.77$\\
MGHD&$2$&$-1523.13$&$0.66$\\
MGHFA&$4$&$-25325.46$&$0.31$\\
\hline
\end{tabular*}}
\label{table:data2comp.}
\end{table}
\begin{table}[!htb]
\centering
\caption{Cross-tabulation of true versus predicted (A,B) classifications from the selected GHD-GLS for the tumour data ($\text{ARI}=0.77$).}
\scalebox{1}{
\begin{tabular*}{0.5\textwidth}{@{\extracolsep{\fill}}lcc}
\noalign{\smallskip}\hline
&A&B\\
\hline
Malignant&352&5\\
Benign&29&183\\
\hline
\end{tabular*}}
\label{table:result2}
\end{table}

\section{Discussion}\label{sec:con}
The GHD-GLS approach for flexible clustering of high-dimensional data was developed based on a mixture of generalized hyperbolic distributions with a penalty term in the likelihood constraining the component-specific concentration matrices. This allows the association structure of the variables to vary across the mixture components. The gamma-lasso penalty used herein enabled the development of an analytically feasible EM algorithm. The BIC with effective number of non-zero parameters was used for model selection. Three simulation studies were carried out to illustrate the proposed GHD-GLS approach and compare with PGMM, MGHD and MGHFA. 
The GHD-GLS approach was also applied to two real datasets and its performance was compared to the MGHD and MGHFA approaches. In the case of the Movehub dataset, the dimensionality was low and the GHD-GLS approach identified three groups, splitting apart the first group that the MGHD approach discovered. Moreover, the GHD-GLS approach was able to find different sparse correlation structures among variables, leading to a simpler interpretation of the clustering results. 
When fitted to the breast cancer diagnostic data, which is often used for benchmarking, the GHD-GLS approach gave superior classification performance when compared to the chosen MGHD and MGHFA. Future work will consider extension of the GHD-GLS approach to the matrix-variate paradigm, where it will be interesting to compare its performance to other approaches \citep[e.g.][]{gallaugher20}.

\section*{Acknowledgements}
This work was supported by a Vanier Canada Graduate Scholarship, the Canada Research Chairs program, an E.W.R. Steacie Memorial Fellowship, and a Dorothy Killam Fellowship.




\appendix

\section{Timing Comparisons for Scenarios 3 and 4}\label{app:time}

\begin{table*}[!ht]
	\centering
	\caption{A comparison of the running times, in seconds, of the selected GHD-GLS, PGMM, MGHD and MGHFA in Scenario~3 (replications=100).}	
	\scalebox{1}{
		\begin{tabular*}{1\textwidth}{@{\extracolsep{\fill}}lccccccccccc}
		  \noalign{\smallskip}\hline
 		&&\multicolumn{2}{c}{GHD-GLS} &\multicolumn{2}{c}{PGMM} &\multicolumn{2}{c}{MGHD} & \multicolumn{2}{c}{MGHFA} \\ 
\cline{3-4}\cline{5-6}\cline{7-8}\cline{9-10}\cline{11-12}
		               &&Time&sd&Time&sd&Time&sd&Time&sd\\
 		 \hline
\multirow{3}{*}{$p=20$}&$n_g=500$ 
& 50.60 &  16.23 & 21.07 & 5.16 & 13.80 & 0.57 &  186.17 &  43.75\\
&$n_g=750$& 45.91 &  26.00  & 33.47 & 8.37 &  21.79 & 0.30 &  271.46  & 65.61\\
&$n_g=1000$& 25.15 & 16.87 &  45.00 & 10.07 & 30.20 & 1.82 & 362.52 & 88.93\\
\hline
\multirow{3}{*}{$p=50$}&$n_g=500$& 122.62 & 59.11  & 56.00 & 6.90 & 40.40 & 1.28 &  377.46 & 87.20 \\
&$n_g=750$&307.93  & 106.36 &89.03  & 9.06 &  62.79 & 2.96 & 554.12 &  139.84\\
&$n_g=1000$& 479.83 & 125.80  & 124.27 & 12.15 & 67.83 & 10.54 & 806.58 & 204.25\\
\hline
\multirow{3}{*}{$p=100$}&$n_g=500$& 134.63 & 68.53 & 203.64  & 22.62 & 72.16 & 0.57 & 1160.90 & 295.96\\
&$n_g=750$& 300.48 & 72.89 &  324.11 &26.33  & 105.78 & 0.81 & 1814.17 &461.93\\
&$n_g=1000$& 638.71 & 120.19  & 442.53 & 36.75 & 140.65 & 1.40 &2800.92 & 811.66 \\
\hline
	\end{tabular*}}
	\label{table:compare3}
\end{table*}
\begin{table*}[!ht]
	\centering
	\caption{A comparison of the running times, in seconds, of the selected GHD-GLS, PGMM, MGHD and MGHFA in Scenario~4 (replications=100).}	
	\scalebox{1}{
		\begin{tabular*}{1\textwidth}{@{\extracolsep{\fill}}lccccccccccc}
		  \noalign{\smallskip}\hline
 		&&\multicolumn{2}{c}{GHD-GLS} &\multicolumn{2}{c}{PGMM} &\multicolumn{2}{c}{MGHD} & \multicolumn{2}{c}{MGHFA} \\ 
\cline{3-4}\cline{5-6}\cline{7-8}\cline{9-10}\cline{11-12}
		               &&Time&sd&Time&sd&Time&sd&Time&sd\\
 		 \hline
\multirow{3}{*}{$p=20$}&$n_g=500$ 
& 65.88 & 42.69  & 4.63 & 2.68 & 15.23 & 0.12  & 101.79 & 3.48\\
&$n_g=750$& 73.46 & 59.01 & 7.79 & 2.49 & 22.29 & 0.11 &151.09  & 4.61\\
&$n_g=1000$&  62.34 & 47.53 & 10.70 &  0.62 & 29.34 &  0.10 & 194.36 & 7.39 \\
\hline
\multirow{3}{*}{$p=50$}&$n_g=500$& 174.17 &  135.97 & 14.60 & 0.75 & 32.46  & 0.15& 218.03 & 13.94\\
&$n_g=750$& 253.29 & 221.71 & 22.98 & 0.42 &  47.85 &0.18  & 326.04  & 13.91 \\
&$n_g=1000$& 353.61 & 327.37 & 31.95 & 0.57 & 63.53 & 0.36 & 431.49 & 24.05 \\
\hline
\multirow{3}{*}{$p=100$}&$n_g=500$& 64.14 & 31.86 & 51.88  & 3.74 &  72.00 & 0.72 & 544.87 & 60.68 \\
&$n_g=750$& 27.27 & 59.15 & 73.81 & 2.62 & 105.66 & 0.92 & 802.73 & 69.51\\
&$n_g=1000$& 18.72 & 17.19  & 100.78 & 1.46 & 139.69 & 1.67 & 1040.10 & 96.63 \\
\hline
	\end{tabular*}}
	\label{table:compare4}
\end{table*}

\end{document}